\pdfoutput=1
\documentclass[sigconf,nonacm]{acmart}

\usepackage[ruled,vlined]{algorithm2e}
\usepackage{subfig}

\AtBeginDocument{%
  \providecommand\BibTeX{{%
    \normalfont B\kern-0.5em{\scshape i\kern-0.25em b}\kern-0.8em\TeX}}}

\begin{document}

\title{DBLog: A Watermark Based Change-Data-Capture Framework}

\author{Andreas Andreakis}
\affiliation{\institution{Netflix}}
\email{aandreakis@netflix.com}

\author{Ioannis Papapanagiotou}
\affiliation{\institution{Netflix}}
\email{ipapapanagiotou@netflix.com}

\renewcommand{\shortauthors}{Andreakis and Papapanagiotou}

\begin{abstract}
It is a commonly observed pattern for applications to utilize multiple heterogeneous databases where each is used to serve a specific need such as storing the canonical form of data or providing advanced search capabilities. For applications it is hence desired to keep multiple databases in sync. We have observed a series of distinct patterns that have tried to solve this problem such as dual-writes and distributed transactions. However, these approaches have limitations with regard to feasibility, robustness, and maintenance. An alternative approach that has recently emerged is to utilize Change-Data-Capture (CDC) in order to capture changed rows from a database’s transaction log and eventually deliver them downstream with low latency. In order to solve the data synchronization problem one also needs to replicate the full state of a database and transaction logs typically do not contain the full history of changes. At the same time, there are use cases that require high availability of the transaction log events so that databases stay as closely in-sync as possible.

To address the above challenges, we developed a novel CDC framework for databases, namely DBLog. DBLog utilizes a watermark based approach that allows us to interleave transaction log events with rows that we directly select from tables to capture the full state. Our solution allows log events to continue progress without stalling while processing selects. Selects can be triggered at any time on all tables, a specific table, or for specific primary keys of a table. DBLog executes selects in chunks and tracks progress, allowing them to pause and resume. The watermark approach does not use locks and has minimum impact on the source. DBLog is currently used in production by tens of microservices at Netflix.
\end{abstract}


\maketitle

\section{Introduction}\label{section:Introduction}
Netfix uses hundreds of microservices performing trillions of operations per day in the data layer. Since there is no single database design that fits all the needs, each of the microservices can utilize multiple heterogeneous databases. For example, a service can use MySQL, PostgreSQL, Aurora or Cassandra for the operational data and Elasticsearch for its indexing capabilities. To be able to keep multiple databases in sync we developed a data enrichment and synchronization platform namely Delta \cite{Delta}. One of the key requirements is to have low propagation delays from the source to the derived stores and that the flow of events is highly available. A key requirement to achieve that is having Change-Data-Capture (CDC) that allows capturing changed rows from a database in near real-time and eventually propagating those rows to downstream consumers \cite{Kleppmann-data-intensive-book}. CDC is becoming increasingly popular for use cases that require keeping multiple heterogeneous databases in sync \cite{kleppmann:onlineeventprocessing, linkedin:databus, facebook_wormhole} and addresses challenges that exist with traditional techniques like dual-writes and distributed transactions \cite{kleppmann:logs}.

In database systems, the transaction log typically has limited retention and it is not guaranteed to contain the full history of changes. Therefore, the full state of a database needs to be captured as well. While operating data synchronization in production at Netflix, we identified some requirements in regards to the full state capture. We wanted to (a) trigger the full state capture at any point in time. That is because the full state may not only be needed initially and may be needed at any time afterwards. For instance if the database is restored from a backup or for repairs if there is data loss or corruption downstream. There are also cases where only a subset of data needs to be repaired, for example if a specific set of rows has been identified to be corrupt downstream. (b) pause or resume at any time so that full state capture does not need to start from the beginning for large tables after restarting the process. (c) capture transaction log events and the full state side by side without stalling one or the other. There are use cases that require high availability of transaction log events so that the replication lag to the source is kept to a minimum. (d) prevent time-travel, by preserving the order of history when transmitting events to a derived datastore. This way an earlier version of a row (like the residential address of a member account) is not delivered after a later version. Hence, a solution had to combine transaction log events and the full state in a way that preserves the history of changes. (e) offer this as a platform. Hence it was crucial to minimize the impact on the source database. Otherwise this can hinter adoption of the platform, especially for use cases that have high traffic. In that regard we want to avoid primitives such as table locks which can block application write traffic. (f) function across a variety of Relational Database Management Systems (RDMBS), such as MySQL, PostgreSQL, Aurora \cite{verbitski2017amazon} etc, that we use in production. In order to achieve that we wanted to avoid using vendor specific features.

\begin{figure}
\centering
\includegraphics[width=0.5\textwidth]{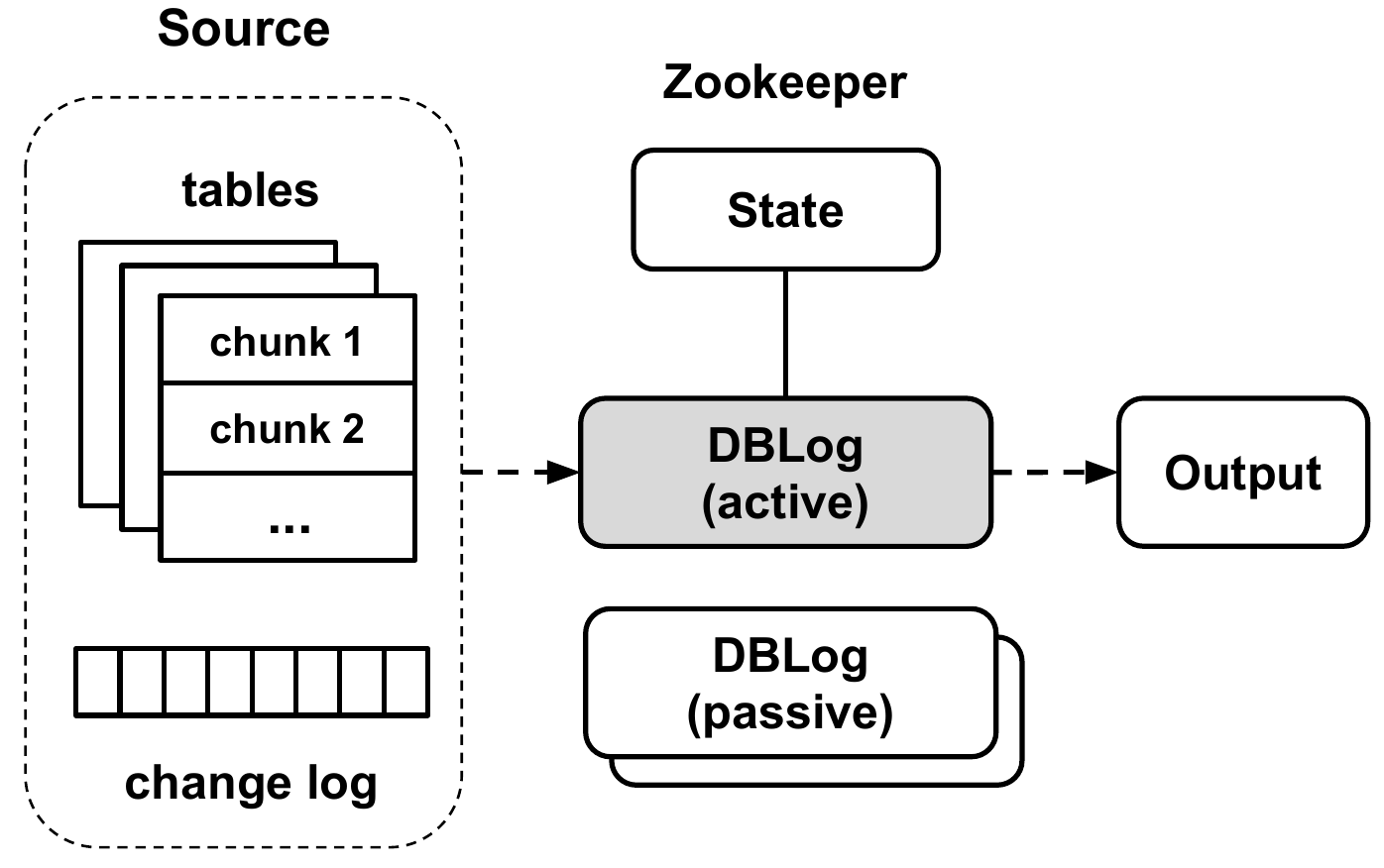}
\caption{DBLog High Level Architecture.}
\label{architecture}
\end{figure}

Based on these requirements we developed DBLog. DBLog runs as a process and utilizes a watermark based approach that allows interleaving of transaction log events with rows that we directly select from tables in order to capture the full state of a database. Our solution allows log events to continue progress without stalling while executing selects. Selects can be triggered at any time on all tables, a specific table, or for specific primary keys of a table. DBLog processes selects in chunks and tracks progress in a state store (currently Zookeeper) allowing them to pause and resume from the last completed chunk. The watermark approach does not use table locks and has therefore minimum impact on the source database. DBLog delivers captured events into an output by using the same format regardless if the origin is the transaction log or a table selection. The output can be a stream like Kafka \cite{kafka} which is a common choice if there is more than one consumer of the events. However DBLog can also write directly to datastores or APIs. DBLog It is also designed with High Availability (HA) in mind by using an active-passive architecture, where one DBLog process is active at a time and multiple passive processes are stand-by and can take over if needed to resume work. Hence, downstream consumers have confidence to receive rows shortly after they changed at the source. Figure \ref{architecture} depicts the high level architecture of DBLog.

\section{Related Work}\label{section:RelatedWork}

We evaluated a series of existing offerings such as: Databus \cite{linkedin:databus}, Debezium \cite{debezium:mysqlsnapshot}, Maxwell \cite{zendesk:maxwell}, MySQLStreamer \cite{yelp:MySQLStreamer}, SpinalTap \cite{airbnb:spinaltap}, and Wormhole \cite{facebook_wormhole}. Existing solutions are similar in regard to capturing events from a transaction log and utilize the same underlying protocols and APIs like MySQL’s binlog replication protocol or PostgreSQL’s replication slots. Captured events are serialized into a proprietary event format and sent to an output which is typically Kafka. Some solutions like SpinalTap and Wormhole only offer log processing without a built-in ability to capture the full state of a database, in which case full state capture needs to be handled out of band. There are existing solutions that have a built-in capability to capture the full state. As a transaction log typically has limited retention it can not be used to reconstruct the full source dataset.  Existing offerings tackle this problem in distinct ways with varying trade-offs:

Databus \cite{linkedin:databus} has a bootstrap service that reads transaction log events from the source and stores them in a separate database. Downstream consumers can access the bootstrap service, if they need to be initialized or for repairs. After bootstrap, consumers start processing log events that originate from the time before the bootstrap so that there is overlap and no events are missed. The catch up from the log can lead to time-travel where row state from the bootstrap may have a more recent row state and an older state is captured from the log afterwards. Eventually the latest state will be discovered from the transaction log.

Debezium \cite{debezium:mysqlsnapshot} captures a consistent snapshot for MySQL and PostgreSQL by using table locks and running selects across all tables within one transaction. Events from the transaction log are then captured from the time after the transaction once all existing rows have been selected. Depending on the implementation and database, the duration of this locking can either be brief or can last throughout the whole selection process, such as with MySQL RDS \cite{debezium:mysqlsnapshot}. In the latter case, write traffic is blocked until all rows have been selected which can be an extended period of time for large databases.

In Maxwell \cite{zendesk:maxwell} a dump is executed by pausing the transaction log processing and then rows are selected from the desired tables. After that, log event processing resumes. This approach is prone to time-travel where a select can return a more recent value of a row and an older value is captured from the log afterwards. Eventually the latest state will be consumed from the log.

MySQLStreamer \cite{yelp:MySQLStreamer} creates a copy of each table at the source, namely a copy table. Then, rows from the original table are inserted into the copy table in chunks resulting into transaction log entries for the inserts. The copy tables are created using the MySQL blackhole engine so that inserts don't occupy table space, while still generating transaction log events. Locking is used to ensure that the order of history is not violated. The MySQLStreamer service then consumes events from the transaction log and is able to detect events that originate from the copy tables, labeling them as events of the original tables. This way downstream consumers receive events per table that either originate from actual application changes or from the copy tables.

Table \ref{tab:1} captures the requirements that we enumerated in section \ref{section:Introduction} for capturing the full state and compares them among existing offerings. We found that no existing approach fulfills the whole range of requirements. Some of the limitations are implied by design such as attempting to select a consistent snapshot first and capturing log events afterwards. The choice of vendor specific features (like the MySQL blackhole engine) is another observed issue, prohibiting code reuse across databases. Some solutions also utilize table locks which can block application write traffic for a short or an extended period of time. Given those observations, we decided to implement a new approach for handling dumps, one that fulfills all our requirements.

\begin{table*}[t]
  \centering
  \begin{tabular}{|l|c|c|c|c|c|}
  \hline
    & Databus\cite{linkedin:databus} & Debezium\cite{debezium:mysqlsnapshot} & Maxwell\cite{zendesk:maxwell} & MySQLStreamer\cite{yelp:MySQLStreamer}& DBLog \\
 \hline
 (a) Can be triggered at any time & \textbf{Yes} & No & \textbf{Yes} & Unknown & \textbf{Yes} \\
 (b) Can be paused and resumed & \textbf{Yes} & No & No & Unknown & \textbf{Yes} \\
 (c) Log event processing does not stall & No & No & No & \textbf{Yes} & \textbf{Yes} \\
 (d) Preserves the order of history & No & \textbf{Yes} & No & \textbf{Yes} & \textbf{Yes} \\
 (e) Does not use locks & \textbf{Yes} &  No & \textbf{Yes} & No & \textbf{Yes} \\
 (f) No vendor specific features & No & No & \textbf{Yes} & No & \textbf{Yes} \\
\hline
\end{tabular}
\caption{Feature Table: Full state capture requirements}
\label{tab:1}
\end{table*}

\section{DBLog}\label{section:DBLog}
DBLog is a Java-based framework, able to capture changed rows from a database’s transaction log and to capture the full state of a database by executing selects on tables. Selects are executed in chunks and are interleaved with log events so that log event processing does not stall for an extended period of time. This is achieved by utilizing a watermark based approach. Selects can be executed at runtime via an API. This allows bootstrapping DBLog’s output with the full state initially or at a later time for repairs. If the output is Kafka with log compaction enabled, then the downstream consumers can be bootstrapped by reading events from Kafka that would contain the full dataset and be continuously updated by appending changed rows as they are captured from the source.  For use cases where there is only one consumer, DBLog can also emit events directly to a datastore or API.

We designed the framework such that the impact to the database is minimal. Selects can be paused and resumed if needed. This is relevant both for failure recovery and to stop processing if the database reached a bottleneck. We also avoided locks on tables so application writes are not blocked. We use Zookeeper \cite{zookeeper} to store progress related to log event processing and chunk selection. We also use Zookeeper for leader election in order to determine the active process while other processes remain idle as passive standbys. We chose Zookeeper because of its maturity, its low latency for reads and writes, its support for linearizable reads \cite{consistency_non_transactional} whenever needed \footnote{Linearizable reads in Zookeeper are provided by calling sync(path) before reading data of the path \cite{zookeeper}\cite{Kleppmann-data-intensive-book}.}, and its availability for writes if a quorum of nodes is reachable. We have built DBLog with pluggability in mind allowing implementations to be swapped as desired, allowing to replace Zookeeper with another datastore.

The following subsections explain transaction log capture and full state capture in more detail.

\subsection{Transaction log capture}
The framework requires a database to emit an event for each changed row in commit order. In MySQL and PostgreSQL a replication protocol exists where the database delivers events shortly after commit time to DBLog via a TCP socket. An event can either be of type: \textit{create}, \textit{update}, or \textit{delete}. For our use cases we assume an event to contain all column values from the time when the operation occurred. Although, DBLog can also be used if a subset of columns is captured. For each event we assume a Log-Sequence-Number (LSN) which is the offset of the event on the transaction log and is encoded as an 8-byte monotonically increasing number. 

Each event is serialized into the DBLog event format and is appended to an output buffer, which is in-memory and part of the DBLog process. Another thread is then consuming events from the output buffer and sends them to the actual output in-order. The output is a simple interface, allowing to plugin any destination, such as a stream, datastore, or generally any kind of service that has an API.

We also capture schema changes. The nature of schema change capture varies among databases, so that there may be schema change deltas in the log, or the database may include the schema information within each emitted event. The way we approach schema capture in DBlog is not covered in this paper due to space limitations.

\subsection{Full state capture}

As transaction logs typically have limited retention they can not be used to reconstruct the full source dataset. When attempting to solve this problem, two of the major challenges are to ensure that log processing does not stall and that the order of history is preserved. One existing solution to this problem is to create a copy of each table at the source database and to populate it in chunks, so that the copied rows will appear in the transaction log in the right order. One can then consume transaction log events and receive the latest state of all rows alongside changed rows \cite{yelp:MySQLStreamer}. This solution however consumes write I/O at the source and requires additional disc space. It is possible to prevent occupying additional table space by using vendor specific features, like the MySQL blackhole engine.

We developed a solution to this problem that only uses commonly available database features and impacts the source database as little as possible. Instead of actually writing the latest state of rows into the transaction log, we are selecting rows from tables in chunks and position the chunks in-memory next to events that we capture from the transaction log. This is done in a way that does preserves the history of log events.

\begin{figure}
\centering
\includegraphics[width=0.3\textwidth]{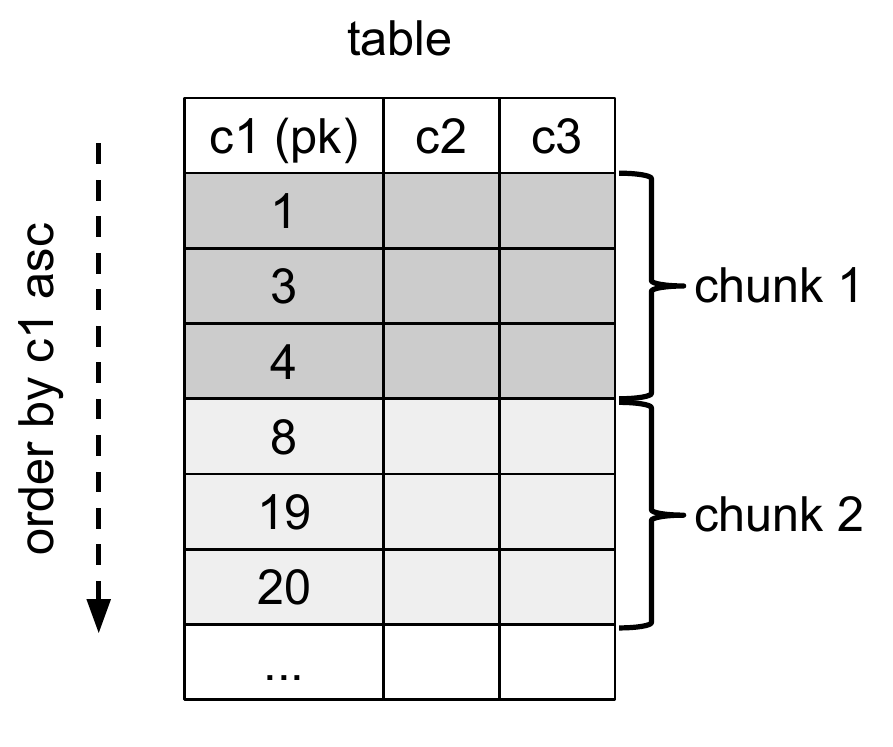}
\caption{Chunking a table with 3 columns c1-c3 and c1 as the primary key (pk). Pk column is of type integer and chunk size is 3. Chunk 2 is selected with the condition c1 $>$ 4.}
\label{Fig:chunking}
\end{figure}

Our solution allows to extract the full state at any time via an API for all tables, a specific table or for specific primary keys of a table. Selects are executed per table and in chunks of a configured size. Chunks are selected by sorting a table in ascending primary key order and including rows, where the primary key is greater than the last primary key of the previous chunk. This query must run efficiently in order to minimize impact on the source database. For these reasons, DBLog requires a database to provide an efficient range scan over primary keys and we only allow selects on tables that have a primary key. Figure \ref{Fig:chunking} is illustrating chunk selection with a simple example.

We are storing the last row of a completed chunk in Zookeeper so that we can pause and resume after the latest completed chunk. Chunks need to be processed in a way that preserves the history of log changes so that a chunk selection that returns an older value can not override newer state that is captured from the transaction log and vice-versa. To achieve this, we create recognizable watermark events in the transaction log so that we can sequence the chunk selection. Watermarks are implemented via a table that we create at the source database. The table is stored in a dedicated namespace so that no collisions occur with application tables. Only a single row is inserted in the table which stores a Universally Unique Identifier (UUID) value. A watermark is then generated by updating the UUID value of that row. The row update results in a change event which is eventually captured by DBLog.

Algorithm \ref{chunk_algorithm} describes the watermark based approach to select the next chunk of a specific table. The algorithm is repeated as long as the table has remaining chunks. Log event processing is briefly paused (step 1). Watermarks are generated by updating the watermark table (steps 2 and 4). The chunk selection occurs between the two watermarks and the chunk is stored in-memory (step 3). After the high watermark is written, we resume log event processing, send received log events to the output, and watch for the low watermark event in the log. Once the low watermark event is received, we start removing rows from the chunk in-memory for all primary keys that changed between the watermarks (step 6). Once the high watermark event is received, we finally append all remaining chunk entries to the output-buffer before processing log events again in a sequential manner (step 7).

\begin{algorithm}
    \caption{Watermark-based Chunk Selection\label{chunk_algorithm}}
    \DontPrintSemicolon
    \KwIn{table}
    \BlankLine
    
    \nlset{(1)} pause log event processing\;
    \BlankLine
    
    lw := uuid(), hw := uuid()\;
    \nlset{(2)} update watermark table set value = lw\;
    \nlset{(3)} chunk := select next chunk from table\;
    \nlset{(4)} update watermark table set value = hw\;
    \BlankLine

    \nlset{(5)} resume log event processing\;
    inwindow := $false$\;
        
    \tcp{other steps of event processing loop}
    \While{true} {
        e := next event from changelog\;
        \uIf{not inwindow}{
            \uIf{e is not watermark}{
                append e to outputbuffer\;
            }
            \uElseIf{e is watermark with value lw} {
                inwindow := $true$\;
            }
        }
       \Else {
            \uIf{e is not watermark}{
                \nlset{(6)} \uIf{chunk contains e.key}{
                    remove e.key from chunk\;
                }
                append e to outputbuffer\;
            }
            \uElseIf{e is watermark with value hw} {
                \nlset{(7)} \For{each row in chunk}{
                    append row to outputbuffer\;
                }
            }
        }
        \BlankLine
        
        \tcp{other steps of event processing loop}
        ...
    }
\end{algorithm}

\begin{figure}
\centering
\subfloat[steps 1-4]{{\includegraphics[width=.47\textwidth]{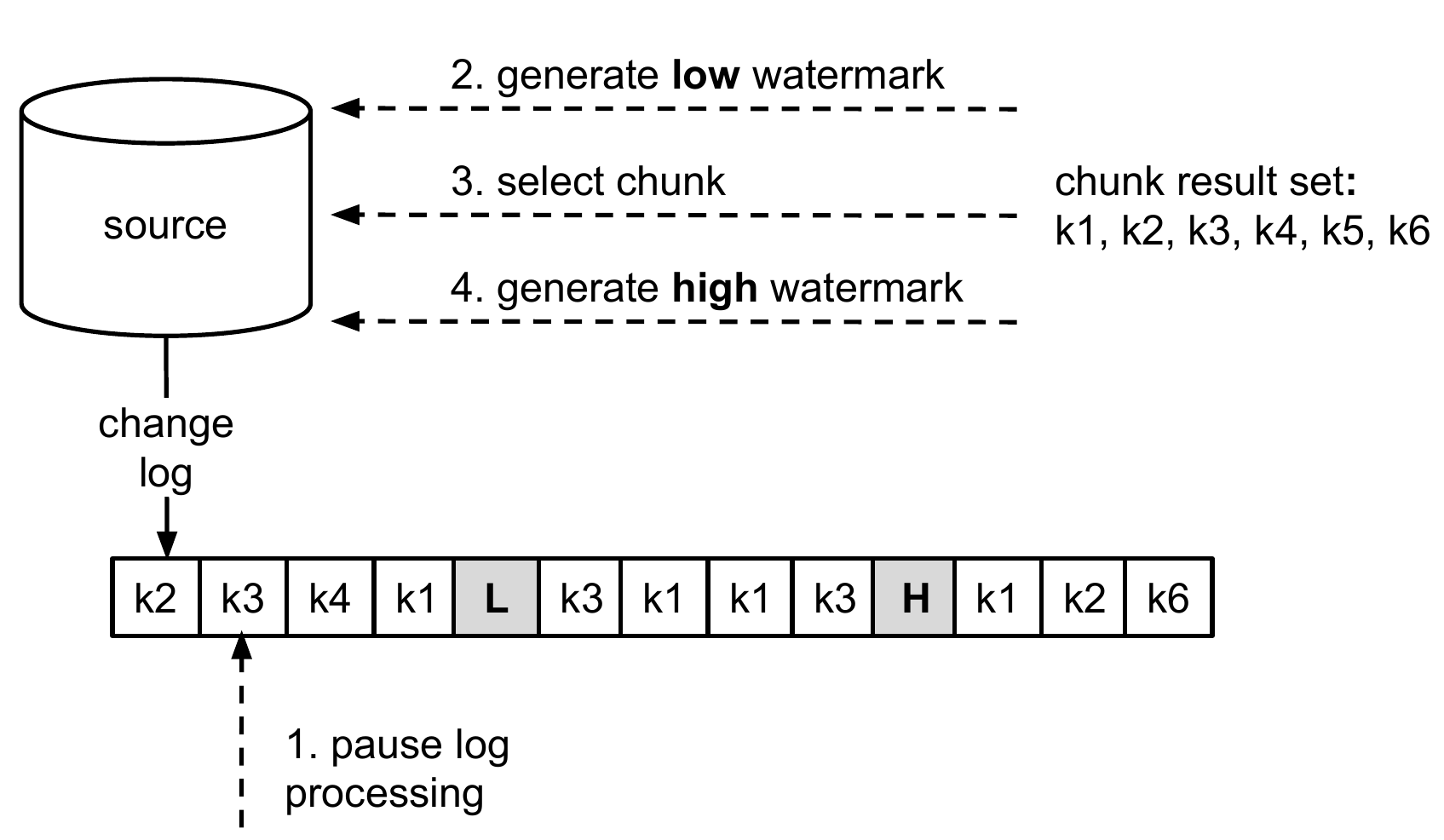}}\label{Fig:watermarkFirstSteps}}
\qquad
\subfloat[steps 5-7]{{\includegraphics[width=.47\textwidth]{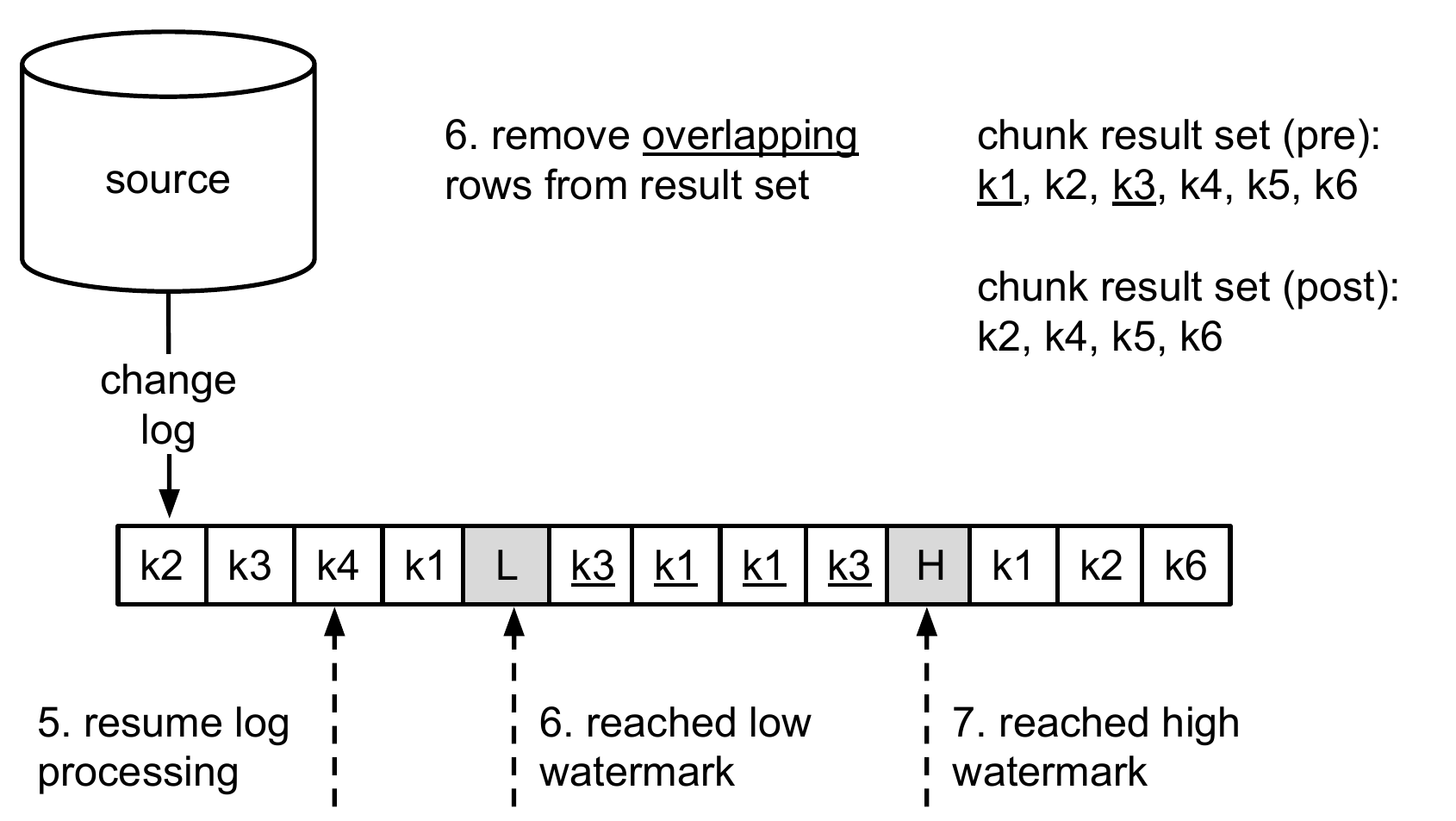}}\label{Fig:watermarkSecondSteps}}
\caption{Watermark-based Chunk Selection}
\end{figure}

The chunk selection of step 3 is required to return state which represents committed changes up to a certain point in history. Or equivalently: the selection executes on a specific position of the transaction log, considering committed transactions up to that point. Databases typically do not expose the execution position of a select on the transaction log (MariaDB being an exception \cite{mariadb_log_position}). The core idea of our approach is to determine a window on the transaction log which guarantees to contain the chunk selection. The window is opened by writing a low watermark, then the selection runs, and the window is closed by writing a high watermark. As the exact position of the select is unknown, all selected chunk rows are removed, which collide with log events within that window. This ensures that the chunk selection can not override the history of log changes. In order for this to work, we must read the table state from the time of the low watermark write, or later (it is fine to include changes that committed after the low watermark write and before the read). More generally, it is required that the chunk selection sees the changes that are committed before its execution. We define this capability as ‘\textit{non-stale reads}’. Additionally, as the high watermark is written afterwards, we require that the select executes before that.

Figures \ref{Fig:watermarkFirstSteps} and \ref{Fig:watermarkSecondSteps} are illustrating the watermark algorithm for chunk selection. We provide an example with a table that has primary keys k1 to k6. Each change log entry represents a create, update, or delete event for a primary key. The steps in the figures correspond to the labels of algorithm \ref{chunk_algorithm}. In figure \ref{Fig:watermarkFirstSteps}, we showcase the watermark generation and chunk selection (steps 1 to 4). Updating the watermark table at step 2 and 4 creates two change events (highlighted with bolt) which are eventually received via the change log. In figure \ref{Fig:watermarkSecondSteps}, we focus on the selected chunk rows that are removed from the result set for primary keys that appear between the watermarks (steps 5 to 7).

Note that a large count of log events may appear between the low and high watermark, if one or more transactions committed a large set of row changes in between. Log event processing is resumed event-by-event after step 4, eventually discovering the watermarks and without ever needing to cache log event entries. Log processing is paused only briefly as steps 2–4 are expected to be fast: watermark updates are single write operations and the chunk selection runs on a primary key index with a limit. Once the high watermark is received at step 7, the non-conflicting chunk rows are appended to the output buffer in-order and ultimately delivered to the output. Appending to the output buffer is a non-blocking operation as the output delivery runs in a separate thread, allowing regular log processing to resume after step 7.

\begin{figure}
\centering
\includegraphics[width=0.45\textwidth]{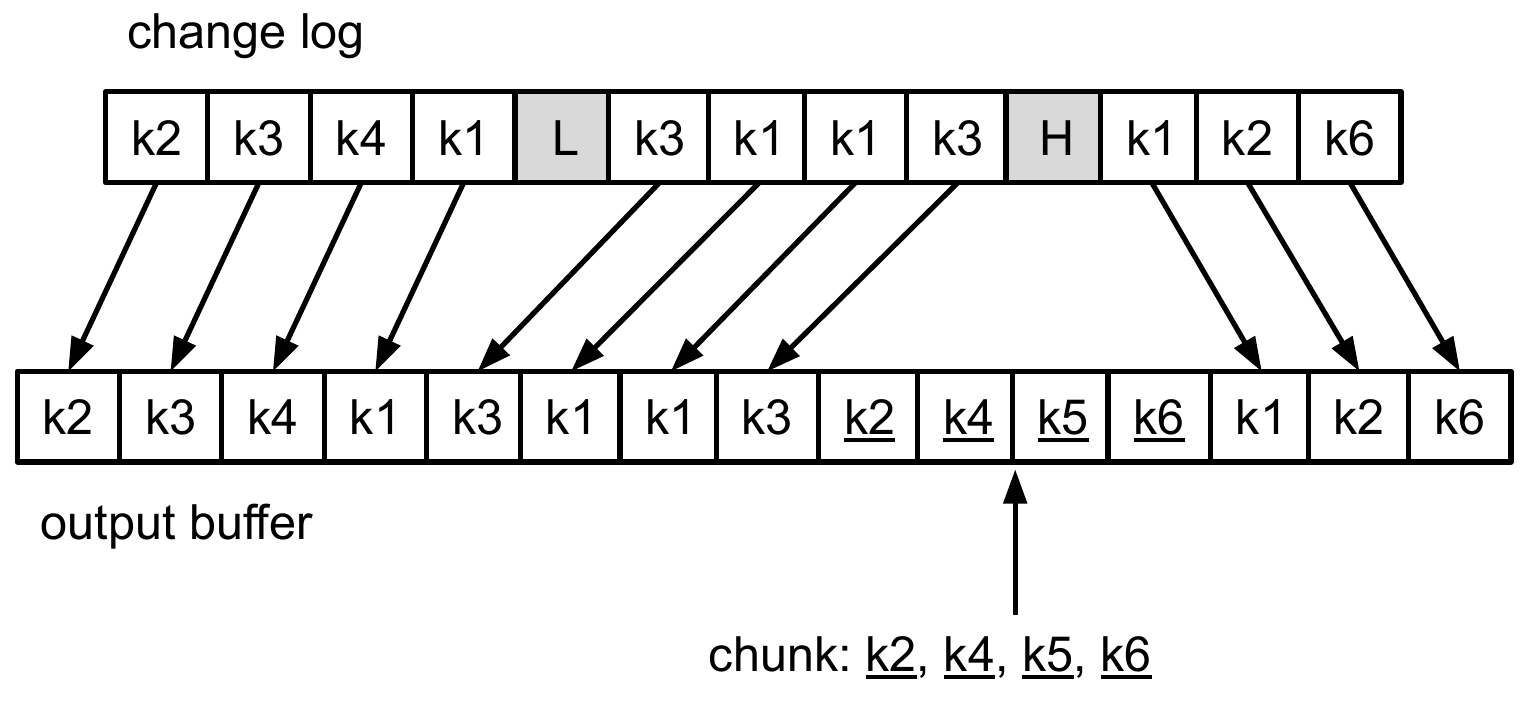}
\caption{Order of output writes. Interleaving log capture with full data capture.}
\label{Fig:Interleave}
\end{figure}

In Figure \ref{Fig:Interleave} we are depicting the order in which events are written to the output, by using the same example as figures \ref{Fig:watermarkFirstSteps} and \ref{Fig:watermarkSecondSteps}. Log events that appear up to the high watermark are appended first. Then, the remaining rows from the chunk selection (underlined entries). And finally, log events that occur after the high watermark. This illustrates the interleaving of log and full data extraction events.

\subsection{Database Support}\label{section:DatabaseSupport}
In order to use DBLog a database needs to emit changed rows from a linear history in commit order and support non-stale reads. These conditions are fulfilled by systems like MySQL, PostgreSQL, MariaDB, etc. so that the framework can be used uniformly across these kinds of databases.

So far, DBLog supports MySQL, PostgreSQL and Aurora. In all cases log events are provided by the database in commit order \cite{mysql_binlog}\cite{postgres_commit_order} and non-stale reads are possible via the read committed isolation level for single select transactions \cite{mysql_consistent_reads}\cite{postgres_read_committed}. To integrate the log events, for MySQL, we use the MySQL Binary Log Connector \cite{mysqlconnector} which implements the binlog replication protocol. For PostgreSQL, we are using replication slots with the wal2json plugin \cite{wal2json}. Changes are received via the streaming replication protocol which is implemented by the PostgreSQL Java Database Connectivity (JDBC) driver. Determining the schema per captured change varies between MySQL and PostgreSQL. In PostgreSQL, wal2json contains the column names and types alongside with the column values. In MySQL schema change deltas are received as binlog events.

Full state capture was integrated by using SQL and JDBC, only requiring to implement the chunk selection and watermark update. The same code is used for MySQL and PostgreSQL and can be used for other databases with JDBC support as well. The dump processing itself has no dependency on SQL or JDBC and allows to integrate databases which fulfill the DBLog framework requirements even if they are not RDBMS databases.

\section{DBLog in Production}\label{section:DBLogInProd}
DBLog is the foundation of the MySQL and PostgreSQL Connectors at Netflix. They are both used in our data synchronization and enrichment platform called Delta \cite{Delta}. DBLog is running in production since 2018 and as of the time this paper is written, it has been deployed in about 30 production use services at Netflix. Those use cases span across heterogeneous data replication, database activity logging, and schema migration.

\textbf{Heterogeneous data replication:} In order to keep track of productions it is crucial to search across all data that are related to movies. This involves data that is managed by separate teams, each of which is owning distinct business entities such as episodes, talents, and deals. These services use MySQL or PostgreSQL in AWS RDS to store their data. DBLog is deployed to each of the involved datastores and captures the full data set and real-time changes into an output stream. The streams are then joined and ingested into a common search index in ElasticSearch, providing search across all involved entities.

\textbf{Database activity logging:} DBLog is also used to log database activity, so that it can be used to inspect what kind of changes occur on the database. In this scenario, changed rows are captured and delivered to a stream. A stream processor is then propagating the events to ElasticSearch (for short-term storage) and Hive (for long-term storage). Kibana is used in ElasticSearch to build activity dashboards so that teams can inspect the amount of occurred operations per table. This is used to inspect data mutation patterns and can be crucial to detect unexpected patterns such drop of inserts to a table after a new service code rolled out with a bug.

\textbf{Schema migration}: When a team is migrating one MySQL database to another
and a new table structure is used in the second database. DBLog is deployed on the old database both to capture the full state as well as new changes as they occur and writes them to a stream. A Flink job is then consuming that data, transforms them to the new table schema format and writes them into the new database. This way, reads for the new database can be verified upfront by running on the populated new schema, while writes still occur to the old schema. In a follow up step, write traffic can also occur to the new schema and traffic on the old database can be stopped.

\section{Conclusions}\label{section:Conclusions}
In this paper, we presented a novel watermark based CDC framework. DBLog capabilities extend the capture of changed rows in real-time from a database transaction log to also extract the full state of a database as part of an integrated offering. In addition, DBLog provides endpoints for users to request the full state and execute it at any time and without stalling log event processing. This is achieved by executing selects on tables in chunks and interleaving the fetched rows with the log events so that both can progress. At the same time, due to the watermark based approach the original order of history is preserved at all times and without using locks on the source database. Moreover, controls are put in place which allow to throttle the chunk selection, or to pause and resume if needed. This is especially relevant when capturing the full state on very large tables and the process crashes, so that the procedure does not need to be repeated from the beginning. DBLog is designed to deliver events to any output, regardless if it is a database, stream, or API. These features open new avenues in synchronizing multiple data systems.

As Netflix operates hundreds of microservices with independent data needs, DBLog has become the foundation of Netflix's data synchronization and enrichment platform. It removed the complexity of application developers in maintaining multiple data stores. DBLog and its watermark based approach is designed to work for RDBMS kind of databases. As a next step, we are working on other CDC frameworks to support databases which don't fall into the DBLog framework, such as multi-master NoSQL databases like Apache Cassandra \cite{cassandra}. The goal is to support similar capabilities as DBLog, namely: the ability to capture the full state at any time, interleave with log events and have minimal impact at the source. 

\begin{acks}
We would like to thank in alphabetic order the following colleagues for contributing to the development of DBLog: Josh Snyder, Raghuram Onti Srinivasan, Tharanga Ga\-mae\-thi\-ge, and Yun Wang.
\end{acks}

\bibliographystyle{ACM-Reference-Format}
\bibliography{dblog}

\end{document}